\documentclass[fleqn,usenatbib]{mnras}

\usepackage[T1]{fontenc}
\usepackage{ae,aecompl}

\usepackage{graphicx}
\usepackage{amsmath}
\usepackage{amssymb}
\usepackage{color}

\usepackage{natbib}
\usepackage{txfonts}
\usepackage{hyperref}
\usepackage{verbatim}
\usepackage{xspace}
\usepackage{breqn}
\usepackage{multicol,lipsum}

\newcommand\rg{GM/c^2}

\newcommand\msun{\rm M_{\rm \odot}}

\newcommand\mdotu{\rm [M_\odot yr^{-1}]}
\newcommand\mbh{M_{\rm BH}}

\title[Polarized radiative transfer with inverse-Compton]{General relativistic polarized radiative transfer with inverse Compton scatterings}

\author[M. Mo{\'s}cibrodzka]{
M. Mo{\'s}cibrodzka$^{1}$\thanks{E-mail: m.moscibrodzka@astro.ru.nl}\\
Department of Astrophysics/IMAPP, Radboud University, P.O. Box 9010, 6500 GL
Nijmegen, The Netherlands
}

\date{Accepted XXX. Received YYY; in original form ZZZ}

\pubyear{2019}

\begin{document}
\label{firstpage}
\pagerange{\pageref{firstpage}--\pageref{lastpage}}
\maketitle

\begin{abstract}
We present {\tt radpol} - a numerical scheme for integrating multifrequency
polarized radiative transfer equations along rays propagating in a curved
spacetime. The scheme includes radiative processes such as synchrotron
emission, absorption, Faraday rotation and conversion, and, for the first
time, relativistic Compton scatterings including effects of light
polarization. The scheme is fully covariant and is applicable to model
radio-$\gamma$-ray emission and its polarization from, e.g., relativistic
jets and accretion flows onto black holes and other exotic objects described
in alternative metric theories and modeled semi-analytically or with
time-dependent magnetohydrodynamical simulations.  We perform a few tests to
validate the implemented numerical algorithms that handle light polarization
in curved spacetime.  We demonstrate application of the scheme to model
broadband emission spectra from a relativistically hot, geometrically thick
coronal-like inflow around a supermassive black hole where the disk model is
realized in a general relativistic magnetohydrodynamical simulation.
\end{abstract}

\begin{keywords}
black hole physics -- MHD -- polarization -- radiative transfer --
relativistic processes
\end{keywords}

\section{Introduction}


Complete models of emission spectra from magnetized plasmas
falling onto a black hole event horizon, or onto a neutron star or ejected as
jets where primary radiative processes are the synchrotron emission and the
inverse-Compton process require incorporating information about light
polarization. The polarized component of light carries additional to
total intensity information about degree of ordering and strength of magnetic
fields, geometry of the relativistic plasma, distribution function of
particles in the plasma and gravitational fields. Including polarization in
the radiative models enable us to investigate in more detaials the physical processes
occurring in a variety of relativistic systems \citep{rees:1975}.

Following pioneering works on strong gravitational effects on polarization
vector
\citep{balazs:1958,connors:1977,stark:1977,connors:1980,ishihara:1988} and
polarized transport through various accretion flows with simplified geometries
(\citealt{angel:1969,laor:1990,haardt:1993,matt:1993,poutanen:1993,agol:1997,agol:2000,dovciak:2004} and many others) recent
focus is on developing numerical fully general relativistic models of polarized transfer
through dynamical plasma models realized in general relativistic magnetohydrodynamical (GRMHD) simulations.

General relativistic {\it polarized} radiative transfer (RT)
schemes are based on ray-tracing approach where RT equations are solved along geodesics lines that connect
emission zone with an observer. Most of the schemes include either synchrotron emission alone
\citep{bromley:2001,broderick:2003,broderick:2004,porth:2011,romans:2012,dexter:2016,moscibrodzka:2018,pihajoki:2018,chan:2019}
or some other form of direct polarized emission \citep{chen:2015}.

An example of fully polarized general relativistic scheme {\tt Pandurata} with
direct synchrotron (or bremsstrahlung) emission and Compton scatterings off
free relativistic electrons coupled to GRMHD simulations has been developed by
\citet{schnittman:2013}. More recently \citet{zhang:2019} developed a code {\tt MONK}
by adding polarization to {\tt grmonty} - a general relativistic transfer
scheme for synchrotron and Compton emission developed by \citet{dolence:2009}
- but modeled polarized emission (due to scattering) in a toy model of a
corona above a thin accretion disk in active galactic nuclei. Both of these
codes use Walker-Penrose constant to parallel transport the polarization
vector along the geodesics \citep{connors:1977} meaning that they are both
limited to Kerr black holes described by Boyer-Lindquist coordinates.  Both
methods also completely neglect the effects of circular polarization, Faraday rotation and conversions effects and
synchrotron emission and self-absorption within the relativistic plasma itself. All
these radiative processes may be negligible for X-rays when these are produced by Compton scattered soft
photons originating in a relativistically cold plasma, but these effects should be incorporated in models in case when
a portion of X-ray emission in a relativistic system is produced by direct
or scattered synchrotron emission, for example, from accelerated electrons or
if one is interested in making connections between emissions in high and low
energies in relativistic systems.

For complete description of multiwavelength polarized emission models in this
work we introduce a numerical scheme that takes into account: fully
general relativistic RT of all Stokes parameters including synchrotron
emission, synchrotron self-absorption, Faraday rotation, Faraday conversion
and Compton scatterings of polarized light off free relativistic electrons. Our scheme is the first one build
within completely covariant framework that will allow one to build radiative models in any
metric and coordinate system in fully three dimensions. Thermal photons from cold plasma can be easily
incorporated in the model, if necessary. Including scattering of photons off bound electrons would require small modifications of the scattering cross-sections and the polarization sensitive scattering kernel. Hence, the new scheme is suitable to model variety of relativistic configurations.

The structure of the article is as follows.  In Sec~\ref{sec:equations}, we
recall equations for polarized RT in relativistic plasmas.  In
Sect.\ref{sec:code}, we describe the scheme developed to solve the polarized
RT in three dimentions. In Sect.~\ref{sec:test1}, we describe and carry out
test of our polarization sensitive Compton scattering scheme and we compare
the numerical results to known semi-analytic solutions.  In
Sect.~\ref{sec:test2} show comparison of the integrations of RT using the new
scheme to results from existing schemes based on a complex plasma
configuration - a hot accretion flow around Kerr black hole realized in
three-dimensional GRMHD simulation.  Sect.~\ref{sec:conclusions} gives a short
summary.

\section{Equations}\label{sec:equations}

\subsection{Photons trajectories}
Photon trajectories are described with null-geodesics equation
\begin{align}\label{eqn:geodesics}
\frac{d^2 x^\mu}{d\lambda^2} = -\Gamma^\mu_{\alpha\beta} \frac{d x^\alpha}{d\lambda} \frac{d x^\beta}{d\lambda}
\end{align}
where $\frac{d x^\mu}{d\lambda} \equiv k^\mu$ is the electromagnetic wave
four-vector, the affine parameter $\lambda$ is defined through
the geodesic equation, and $\Gamma$'s are the connection
coefficients that account for curvature of a stationary metric. The metric itself is
a parameter of a model. 

\subsection{Synchrotron processes}\label{sec:synchrotron}
We consider polarized radiative transfer along geodesics through relativistic, fully ionized, magnetized
plasma in which the dominating radiative transfer processes are synchrotron
emission/absorption, Faraday effects and inverse-Compton process. The
equations for evolution of Stokes parameters due to synchrotron
emission/self-absorption and Faraday effects is:
\begin{equation}\label{spoltrans}
\frac{d}{d\lambda}
\begin{pmatrix} I \\ Q \\ U \\ V \end{pmatrix}
 = \begin{pmatrix} j_{I} \\ j_{Q} \\ j_{U} \\ j_{V} \end{pmatrix}
- \begin{pmatrix}
\alpha_{I} & \alpha_{Q} & \alpha_{U} & \alpha_{V} \\
\alpha_{Q} & \alpha_{I} & \rho_{V} & -\rho_{U} \\
\alpha_{U} & -\rho_{V} & \alpha_{I} & \rho_{Q} \\
\alpha_{V} & \rho_{U} & -\rho_{Q} & \alpha_{I}
\end{pmatrix}
\begin{pmatrix} I \\ Q \\ U \\ V \end{pmatrix} 
\end{equation}
where $j_{IQUV}$ is the synchrotron emissivity, $\alpha_{IQUV}$ is the absorptivity and $\rho_{QUV}$ is the Faraday rotation/conversion,
and where the absence of subscript $\nu$ implies that a term appears in
invariant form, i.e. $j_I = j_{\nu,I}/\nu^2$, $\alpha_Q = \nu \alpha_{\nu,Q}$,
and $\rho_V = \nu \rho_{\nu,V}$ and the derivative
is understood to follow an individual photon in frequency space.  
Notice that Eq.~\ref{spoltrans} is written down in the rest-frame that is co-moving with
the plasma, in which transfer coefficients are defined, and it does not describe how to parallel transport the polarization vector.  
The covariant formulation of polarization vector transport is presented, e.g., in \citet{gammie:2012} who re-wrote relativistic transport equation in terms of coherency tensor $N^{\alpha\beta}$:
\begin{equation}\label{eq:trans1}
k^\mu \nabla_\mu N^{\alpha\beta} = J^{\alpha\beta} + H^{\alpha \beta \gamma \delta} N_{\gamma \delta} 
\end{equation}
where $\nabla_\mu$ is a covariant derivative, $J^{\alpha\beta}$ includes the emissivity coefficients, and $H^{\alpha \beta \gamma \delta}$ incorporates absorption and Faraday coefficients. The transformation of tensor components to Stokes parameters and vice-versa is trivial and requires defining an orthonormal tetrad $e^\mu_{(a)}$. The method of solving Eqs~\ref{spoltrans} and~\ref{eq:trans1}, its numerical implementation ({\tt ipole} code) with extensive testings and comparison to independent scheme \citep{dexter:2016} and application to astrophysical problem is presented in our recent publication \citep{moscibrodzka:2018}. 

\subsection{Compton scattering}

We consider Compton scatterings of synchrotron photons on free relativistic
electrons (Synchrotron Self-Compton emission). The Compton scattering event is
described in a tetrad in which electron is at rest, i.e.,
$e^\mu_{(0)}=p^\mu_e$ where $p^\mu_e$ is the four-momentum of an electron
measured in the fluid co-moving frame. The z-axis if chosen to be aligned with
the spatial part of the wavevector of the incident photon $e^\mu_{3}=k^\mu - \epsilon_e p_e^\mu$.
The x and y axes ($e^\mu_1$ and $e^\mu_2$) are chosen by orthogonalization. 

The differential Klein-Nishina (KN) cross section for scattering of polarized photon on an unpolarized electron is (\citealt{fano:1949}, \citealt{bere:1982}): 
\begin{equation}\label{eq:KN}
\frac{d\sigma^{KN}}{d\Omega} = \frac{1}{4} r_e^2
\left(\frac{\epsilon_e'}{\epsilon_e}\right)^2 \left[F_0 
+  F_{11} \xi_1 \xi_1^' + F_1 (\xi_1  + \xi_1^')
+  F_{22} \xi_2 \xi_2^' + F_{33} \xi_3 \xi_3^'  \right]
\end{equation}
where $r_e$ is the electron classical radius, $\epsilon_e$ and $\epsilon_e'$ are incident and
scattered energy of photon in units of $m_ec^2$,
$\xi_{1,2,3}$ ($\xi_{1,2,3}'$) are normalized polarizations of incident (scattered) photon, i.e.,
$\xi_1=Q/I$, $\xi_2=U/I$, and $\xi_3=V/I$, and $F$ coefficients are elements
of the Fano matrix (\citealt{fano:1949,fano:1957})\footnote{Notice that \citealt{bere:1982} (and many others) use different
index numbers for fractional Stokes parameters and for the Fano matrix elements. Here we use
indexing that is consisitent with notation used in Sect.~\ref{sec:synchrotron}.}:
\begin{align}
\label{eq:fano}
\begin{split}
  F_0=\frac{\epsilon_e'}{\epsilon_e} + \frac{\epsilon_e}{\epsilon_e'} - \sin^2\theta'\\
  F_1=\sin^2\theta'\\
  F_{11}=1+\cos^2\theta'\\
  F_{22}=2 \cos\theta'\\
  F_{33}=\left( \frac{\epsilon_e'}{\epsilon_e} + \frac{\epsilon_e}{\epsilon_e'} \right) \cos \theta'
\end{split}
\end{align}
where $\theta'$ is the polar angle of scattering and $d\Omega=\sin \theta' d\theta' d\phi'$. 
The dependency of the cross section on the azimuthal angle $\phi'$ is implicit since parameters $\xi_{123}$ are defined in a coordinate system that is fixed to the scattering plane (plane normal to ${\bf k}\times {\bf k'}$).
The transformation of fractional polarizations in a scattering event is given by:
\begin{align}\label{eq:fractions}
  \begin{split}
    \xi_1'=\frac{F_1 + \xi_1 F_{11}}{F_0+\xi_1 F_1}\\
    \xi_2'=\frac{\xi_2 F_{22}}{F_0+\xi_1 F_1}\\
    \xi_3'=\frac{\xi_3 F_{33}}{F_0+\xi_1 F_1}.
  \end{split}
\end{align}
and the transformation of the Stokes vector is described by:
\begin{equation}\label{eq:fano_matrix}
\begin{pmatrix} I' \\ Q' \\ U' \\ V' \end{pmatrix}
 = \left(\frac{\epsilon_e'}{\epsilon_e}\right)^2 \begin{pmatrix}
   F_0 & F_1 & 0 & 0\\
   F_1 & F_{11} & 0 & 0\\
   0 & 0 & F_{22} & 0\\
   0 & 0 & 0 & F_{33}
\end{pmatrix}
\begin{pmatrix} I \\ Q\\ U \\ V \end{pmatrix}
\end{equation}
where the primed Stokes are Stokes parameters after scattering.

In Eq.~\ref{eq:KN} taking sum over all possible polarizations of the scattered photon $\xi_{123}'$ one can obtain the total cross section as a function of incident photon polarization:
\begin{equation}\label{eq:KN3}
\frac{d\sigma^{KN} (\xi_{123}) }{d\Omega} = \frac{1}{2} r_e^2 \left( \frac{\epsilon_e}{\epsilon_e'} + \frac{\epsilon_e'}{\epsilon_e} - (1-\xi_1) \sin^2\theta' \right).
\end{equation}
Eq.~\ref{eq:KN3} depends on linear polarization but is independent of Stokes U and V. For unpolarized incident photon and for elastic scattering ($\epsilon_e'=\epsilon_e$) Eq.~\ref{eq:KN3} reduces to classical Thomson scattering cross section and Eq.~\ref{eq:fano} becomes Rayleigh scattering phase matrix.

Notice that the total KN cross section for incident polarized light (integrated over all
scattering angles $\theta'$ and $\phi'$) is independent of polarization:
\begin{dmath}\label{eq:KNtot}
\sigma^{KN}= \int \frac{d\sigma^{KN}}{d\Omega} d\Omega = \sigma^{TH} \frac{3}{4\epsilon_e^2} \left( 2 + \frac{\epsilon_e^2(1+\epsilon_e)}{(1+2\epsilon_e)^2} + \frac{\epsilon_e^2-2\epsilon_e-2}{2\epsilon_e} \log(1+2\epsilon_e) \right).
\end{dmath}

\section{Numerical method}\label{sec:code}

Our numerical methods to solve above equations extends existing
unpolarized radiative transfer scheme {\tt grmonty}, described in
detail by \citet{dolence:2009}.  {\tt grmonty} is a Monte Carlo general
relativistic scheme simulating  synchrotron emission and
self-synchrotron and inverse-Compton scatterings. The code generates mock
multiwavelength spectra for chosen plasma configuration and chosen metric
tensor. Here we modify {\tt grmonty} to also track light polarization along their
original transport equations and to investigate how including polarization of
light modifies the RT solutions. In what follows, we describe additions to the
original scheme that allow us to follow polarization of synchrotron photons
and polarization in relativistic Compton scatterings. Our new numerical scheme
is referred as to {\tt radpol}. 

\subsection{Polarized photons in Monte Carlo scheme: sampling and weights}\label{sec:tetrads}

The radiation field is represented by ``superphotons''. Each superphoton is described by three quantities:
coordinate $x^\mu$, wave four-vector $k^\mu$ and  weight $w$.
The weight $w$ is a scalar proportional to the number of photons in a single
superphoton. To generate superphotons we follow
exactly scheme described in Section 2 in \cite{dolence:2009}. A
modification is made to assign superphoton polarization state: namely each superphoton has now
assigned four weights $w_{\rm I}$, $w_{\rm Q}$, $w_{\rm U}$,
$w_{\rm V}$. Weight $w_{\rm I}$ (initially $w_{\rm I} \equiv w$)
is invariant under Lorentz transformations and it is
proportional to the invariant radiation specific intensity ($I_{\rm inv}=I/\nu^3$ where $\nu$ is
photon frequency). The three remaining weights are proportional to the
``invariant'' Stokes parameters:
\begin{equation}
w_{\rm Q} \propto Q_{\rm inv}\equiv Q/\nu^3,\\ 
w_{\rm U} \propto U_{\rm inv}\equiv U/\nu^3,\\
w_{\rm V} \propto V_{\rm inv}\equiv V/\nu^3.
\end{equation}
While $w_I$ is positive definite, $w_{\rm Q}$,
$w_{\rm U}$, $w_{\rm V}$ can be negative so they do not have the same physical
meaning as $w_I$, but they only quantify polarization of a superphoton.  

Each superphoton $k^\mu$ and $w_{\rm I}$ are selected using rejection sampling
based on plasma synchrotron
emissivity for Stokes I, $j_I$. The remaining weights are defined using local
synchrotron emissivities for Stokes $Q$, $U$ and $V$, i.e., $w_Q=w_I j_Q/j_I$,
$w_U=0$, and $w_V=w_I j_V/j_I$. The initial superphoton wavevector $k^\mu$ and the 
polarization vector ${\bf w}_S=(w_I,w_Q,w_U,w_V)$ are defined in two different
specific reference frames (tetrads).

The wavevector $k^{(a)}$ is sampled in an ortho-normal tetrad attached to the
fluid element (plasma frame). The first tetrad basis vectors are $e^\mu_{(0)}=u^\mu$ and
$e^\mu_{(3)}=b^\mu$, where $b^\mu$ is the magnetic field vector measured in
the fluid frame. The other two basis vectors $e^\mu_{(1)}$ and $e^\mu_{(2)}$ are chosen via
Gram-Schmidt orthonormalization procedure. The tetrad-to-coordinate basis
Lorentz transformation for the superphoton wavevector is $k^\mu=e^\mu_{(a)}
k^{(a)}$.

A different ortho-normal tetrad is built to tetrad-to-coordinate basis
transformation for the polarization wavevector. This is because the
synchrotron emissivities for Stokes parameters
are defined in a special frame in which $j_{\rm U}=0$. This tetrad
is defined as: $\tilde{e}^\mu_{(0)}=u^\mu$ and 
$\tilde{e}^\mu_{(3)}=k^\mu - \omega u^\mu$, where $\omega=-k^\mu u_\mu$ is the
superphoton frequency measured by the observer with four-velocity $u^\mu$. 
Notice, the third spatial basis element is a unit vector oriented parallel to
the spatial component of the wavevector.
Two other basis vectors are chosen via Gram-Schmidt orthonormalization procedure.

\subsection{Parallel transport, absorption, Faraday rotation and conversion of a superphoton}

$N$ synchrotron superphotons is sampled and launched from 
the magnetized plasma. The plasma synchrotron emissivity is accounted for by assigning to each
superphoton an initial polarization weights ${\bf w}_S=(w_I,w_Q,w_U,w_V)$. The
weights are assigned in the plasma frame and translated into coherency tensor $N^{\alpha\beta}$
using the fluid tetrad. 
The parallel transport of polarization weights is then carried out by solving modified
Eq.~\ref{eq:trans1}, i.e., leaving out all terms for plasma
emissivity in transport equations:
\begin{equation}\label{eq:trans2}
k^\mu \nabla_\mu N^{\alpha\beta} = H^{\alpha \beta \gamma \delta} N_{\gamma \delta}
\end{equation}
where coherency tensor $N^{\alpha\beta}$ now contains Stokes weights (instead of Stokes parameters.)\footnote{$N^{\alpha\beta}$ relates to the Stokes weights ${\bf w}_S$ is a same way as it relates to the standard Stokes parameters} and
the right-hand side of Eq.~\ref{eq:trans2} accounts for self-absorption and Faraday effects only. 
The numerical method used to integrate Eq.~\ref{eq:trans2} is adopted from
{\tt ipole} scheme \citep{moscibrodzka:2018} who applied second-order split
scheme for parallel transport plus source step to evolve $N^{\alpha\beta}$.

In the source step, we read-out ${\bf w}_S$ from $N^{\alpha\beta}$ in the
fluid frame and we solve (leaving out the emissivity coefficients accordingly):
\begin{equation}\label{eq:spoltrans2}
\frac{d}{d\lambda}
\begin{pmatrix} w_I \\ w_Q \\ w_U \\ w_V \end{pmatrix}
 = - \begin{pmatrix}
\alpha_{I} & \alpha_{Q} & \alpha_{U} & \alpha_{V} \\
\alpha_{Q} & \alpha_{I} & \rho_{V} & -\rho_{U} \\
\alpha_{U} & -\rho_{V} & \alpha_{I} & \rho_{Q} \\
\alpha_{V} & \rho_{U} & -\rho_{Q} & \alpha_{I}
\end{pmatrix}
\begin{pmatrix} w_I \\ w_Q \\ w_U \\ w_V \end{pmatrix} 
\end{equation}
Following \citet{moscibrodzka:2018} arguments we assume that transfer coefficients in
Eq.~\ref{eq:spoltrans2} are constant along a single step on geodesics. 
In such case, Eq.~\ref{eq:spoltrans2} has simple solution:
\begin{equation}\label{eq:sol}
\begin{pmatrix} w_I \\ w_Q \\ w_U \\ w_V \end{pmatrix} (\lambda)
 = O_{AB}(\lambda-\lambda_0)
\begin{pmatrix} w_I \\ w_Q \\ w_U \\ w_V \end{pmatrix} (\lambda_0)
\end{equation}
where analytic expression for operator $O_{AB}(\lambda-\lambda_0)$ has been found by
\citet{LDI:1985}. Also notice that operator $O_{AB}$ is dimensionless so
that ${\bf w}_S$ can have arbitrary units. 
For unpolarized superphotons, i.e., $w_{QUV}=0$, the Eq.~\ref{eq:sol} becomes:
\begin{equation}
\frac{dw_I}{d\tau_a}=-w_I,
\end{equation}
where
\begin{equation}
d\tau_a= \alpha_{I} {\mathcal C} d\lambda 
\end{equation}
and ${\mathcal C}\equiv Lh/m_ec^2$ is a constant that depends on units of
$\lambda$ and $k^\mu$. Quantity $L$ is the lengthscale associated with a given
physical problem, e.g., for RT around black hole, $L\equiv \rg$.
In the limit of unpolarized transport our new method is 
consistent with solutions originally implemented in {\tt grmonty} because
neither the parallel transport step nor the Lorentz transformations from
coordinate frame to plasma frame (and vice versa) affect superphoton
weight $w_I$ ($w_I$ is a scalar proportional to the number of
photon quanta in a given superphoton).

\subsection{Scattering of polarized radiation on relativistic electrons}

When transporting light along geodesics an optical depth for Compton
scattering, $\tau_{sc}$, is computed. The optical depth for scattering is defined as
\begin{equation}
d\tau_{sc} = \alpha_{sc} {\mathcal C} d\lambda
\end{equation}
where $\alpha_{sc}$ is the Lorentz invariant scattering opacity given by
integrating total KN cross section (Eq.\ref{eq:KNtot})
over the assumed electron distribution function.
In the event of scattering the wavevector $k^\mu$ and coherence tensor $N^{\alpha\beta}$ are Lorentz boosted into the plasma frame where for a given $k_{plasma}^\mu$ an electron energy and momentum is sampled from an isotropic electron distribution function. Given electron four-momentum in the plasma frame we construct another tetrad, defined within the plasma tetrad, in which the electron is at rest ($e_t^\mu=p_e^\mu$), the incident photon wavevector is aligned with spatial component of the third tetrad basis vector $e_3^\mu=k^\mu_{plasma}-\epsilon_e p_e^\mu$ where $\epsilon_e=-k_{plasma}^\mu p_{e,\mu}$ is the incident photon energy measured in the electron rest-frame.
Basis vector $e_2^\mu$ and $e_1^\mu$ are chosen via orthonormalization. Two angles describe the scattered photon beam $\theta' \in (0,\pi)$ and $\phi' \in (0,2\pi)$.
Basis vector $e_2^\mu$ defines azimuthal scattering angle $\phi'=\pi/2$ and if the photon gets scattered along $e_3^\mu$ - then the polar scattering angle $\theta'=0$.

We simulate Compton scattering in the electron frame as follows.
1)  We sample the cosine of polar angle $\cos(\theta')$ and azimuthal $\phi'$ angle of
  scattered superphoton from uniform distributions. Azimuthal angle defines the scattering plane. 2) We
rotate ${\bf w}_S$ with respect to that azimuthal angle and calculate
$\xi_1=w_Q/w_I$.
3) The energy of scattered superphoton $\epsilon_e'$ is computed from kinematic
relation:
$\epsilon_e'=1/(1-\cos\theta+1/\epsilon_e)$.
In case of incident photon with small energy in comparison to electron rest
mass energy ($\epsilon_e \ll 1$) the scattering becomes elastic independently of scattering angle.
4) Given $\epsilon_e$, $\epsilon_e'$, and scattering angles $\phi'$ and
$\theta'$, the transformation of a polarization vector ${\bf w}_S$ in done
using Eq.~\ref{eq:fano_matrix} (notice that we do alter the $w_I$ in a
scattering event because the scattering angles are randomly chosen rather than
drawn from the distribution function).
5) Given the new ${\bf w}_S'$ we build a new coherency tensor of scattered photon that is transformed from electron to fluid and then from fluid to coordinate frame for further parallel transport with the scattered wavevector.
6) In the final step of the scattering simulation, a wavevector of scattered beam is constructed with:
\begin{align}
\begin{split}
k^{(0)}=\epsilon_e'\\
k^{(1)}=\epsilon_e' \sin\theta' \cos\phi'  \\
k^{(2)}=\epsilon_e' \sin\theta' \sin\phi' \\
k^{(3)}=\epsilon_e' \cos\theta' \\
\end{split}
\end{align}
which is transformed from electron rest-frame to fluid frame and then from fluid frame to coordinate frame where it is parallel transported along geodesic equation.

\subsection{Polarized spectrum recorded by an external observer}

To construct observed polarized spectrum of the plasma model, we build a polarimetric
``detector''. The detector is another tetrad defined as:
$e^\mu_{(0)}=u^\mu_{d}$ where $u^\mu_{d}$ is the detector four-velocity defined by user (by default we set a stationary observer). 
The detector is located at a distance $r_{d}$ from the origin
of the utilized coordinate system. 
The detector occupies solid angle $d\Omega \equiv d \phi_d 
d \theta_d \sin(\theta_d)$, 
where the angular size is given by $\Delta \phi_{d}$ and $\Delta \theta_{d}$
and detector it is centered at $(\theta_{d,0},\phi_{d,0})$, which are free parameters.

Measuring polarization of a superphoton requires projection of the $N^{\alpha\beta}$
into the detector tetrad and extraction of weights for each Stokes parameter. 
Our polarimetric detector has energy bins from radio
to $\gamma-$ray frequencies. All sampled superphotons are independent of one
another, hence in case of multiple superphotons falling into the same energy
bin, their Stokes parameters (or weights in this case) can be summed over \citep{chandra:1960}.
We translate ${\bf w}_S$ measured in the detector frame into luminosities using
following relations:
\begin{align}
\label{eq:L}
\begin{split}
\nu L_{I,\nu,i} = \frac{4\pi}{\Delta \Omega \Delta t} \frac{1}{\Delta \ln
  \nu}\sum_j w_{I,j} h\nu_j, \\
\nu L_{Q,\nu,i} = \frac{4\pi}{\Delta \Omega \Delta t} \frac{1}{\Delta \ln
  \nu}\sum_j w_{Q,j} h\nu_j, \\
\nu L_{U,\nu,i} = \frac{4\pi}{\Delta \Omega \Delta t} \frac{1}{\Delta \ln
  \nu}\sum_j w_{U,j} h\nu_j, \\
\nu L_{V,\nu,i} = \frac{4\pi}{\Delta \Omega \Delta t} \frac{1}{\Delta \ln \nu}\sum_j w_{V,j} h\nu_j.
\end{split}
\end{align}

\section{Simple test problems}\label{sec:test1}

We carry out two simple tests to validate the polarization sensitive scattering kernel. Both tests assume Minkowski spacetime. 

\subsection{Scattering of unpolarized beam off one relativistic electron}

\begin{figure*}
\centering
\begin{center}
\includegraphics[width=0.48\textwidth,angle=0]{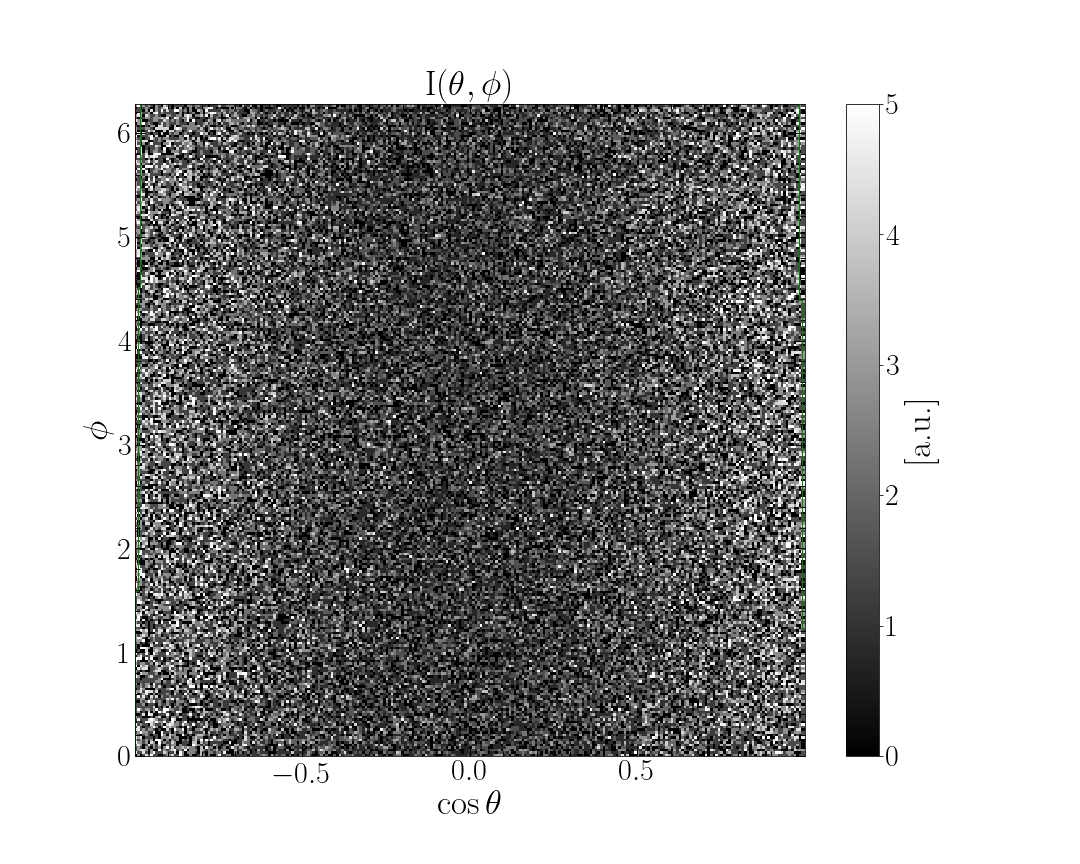}
\includegraphics[width=0.48\textwidth,angle=0]{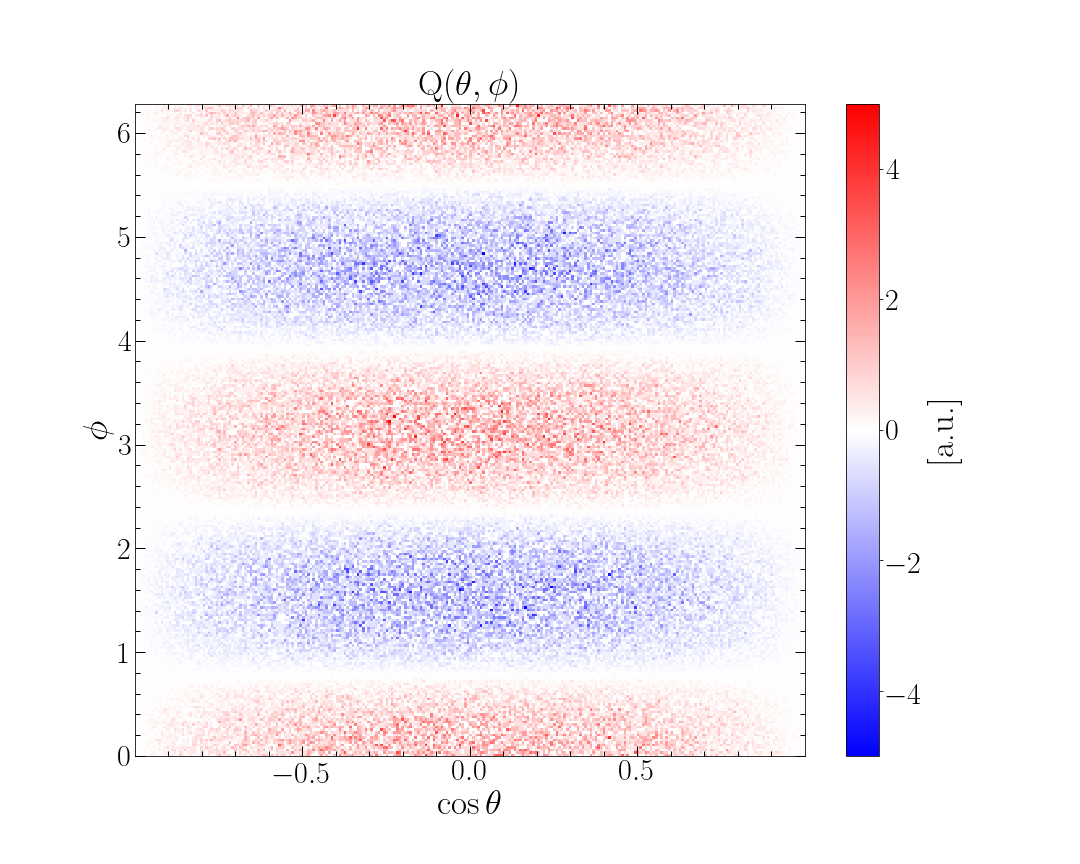}
\caption{Stokes I and Q of Thomson scattered initially unpolarized light off electron
  at rest as a function of the scattering polar and azimuthal angle. The
  Stokes parameters have arbitrary units.}\label{fig:compton_simple}
\end{center}
\end{figure*}

In the first test we consider a monochromatic, unpolarized
beam of light scattering off a single electron with a fixed
$p_e^\mu=(1,0,0,0)$. Here we consider scattering in Thomson regime.
Fig.~\ref{fig:compton_simple} shows maps of Stokes I and Stokes Q of
scattered light as a function of scattering angle $\theta'$ and $\phi'$.
The numerical calculations agree with
theoretical prediction, scattered light on electron at rest is more linearly polarized
when scattered at 90 degrees and the polarization angle depends on $\phi'$ angle. For very small and close to maximum scattering polar angles the scattered light remains unpolarized. 

\subsection{Scattering of polarized beam off a population of relativistic electrons}

In the second test we consider a monochromatic, polarized
beam of light scattering off a population of relativistic electrons described
by Maxwell-J{\"u}ttner distribution function parameterized by dimensionless
electron temperature $\Theta_e \equiv k_BT_e/m_ec^2$. The analytic solutions
to this problem has been found by \citet{bonometto:1970}. We describe these solutions and their limitations in detail in the appendix.
In Fig.~\ref{fig:bonometto} we show agreement between the
numerical solutions and the analytical expectations for thermal electrons with
$\Theta_e=100$ and various incident beam linear and circular
polarizations. We show results for light scattered in direction $(\theta',\phi')=(85^\circ,0)$.

\section{Complex test problems}\label{sec:test2}

\subsection{Astrophysical problem}\label{sec:test_setup}

Our numerical radiative transfer scheme inherits many of the numerical methods
from {\tt grmonty} and {\tt ipole} codes, both of which have been extensively tested
assuming simple and complex plasma conditions in flat, twisted and curved spacetimes
(see e.g. \citealt{moscibrodzka:2018}, \citealt{moscibrodzka:2019}). The main application of the {\tt radpol} scheme
is modeling multiwavelength emission from relativistic jets and accretion flows 
around compact objects. Here we test the performance of {\tt radpol}
and compare it to it's "parent" codes using a model of complex
accretion flow. The background stationary spacetime is described by Kerr
metric parameterized with dimensionless spin parameter $a_*\equiv Jc/GM^2$
with |$a_*<1$| (where $M$ and $J$ are, respectively, the mass and the angular
momentum of a black hole) and Kerr-Schild coordinate system.
The underlying magnetized plasma dynamics
is simulated using 3D GRMHD code. We consider radiative transfer through a
geometrically thick, relativistically hot disk accreting onto spinning
($a_*=0.9375$) black hole \citep{gammie:2003}.

To construct mock observations of a simulation, one has to specify the free
parameters of the RT model: the central black hole mass $\mbh$, accretion rate
$\dot{M}$ (which scales the dimensionless density and magnetic field strength), electron model (here we assume strong coupling between heavy ions and electrons and that electrons are described by the Maxwell-J{\"u}ttner distribution function) and the observer's viewing angle $i$ measured with respect to the coordinate $\theta=0$ direction. We assume $\mbh=6.5 \times 10^9 \msun$, $\dot{M}=4 \times 10^{-4} \mdotu$ and $i=60^\circ$ (see \citealt{moscibrodzka:2016,moscibrodzka:2017} for details). These parameters are close to ones used to model emission from
the core of M87 galaxy. However, here we do not attempt to
model any particular observations, but only demonstrate the scheme performance.

\subsection{Comparison to {\tt ipole}: polarized synchrotron spectra}

{\tt ipole} is an imaging ray tracing code 
in which radiative transfer equations are solved along
null-geodesics that terminate at a ``detector'' at some large distance
from the central object where a polarization map at a chosen observing
frequency is constructed. In {\tt ipole}
simulations the synchrotron emission, synchrotron self-absorption and Faraday
rotation and conversion are included but the inverse-Compton scatterings
are not accounted for. Hence we can check consistency of {\tt radpol} and {\tt
  ipole} in the direct synchrotron emission only. {\tt ipole} is producing
images at a given frequency, to construct spectra we generate frames for many
frequencies and integrate intensity of all Stokes parameters over each polarimetric map.

\begin{figure*}
\centering
\begin{center}
  \includegraphics[width=0.48\textwidth,angle=0]{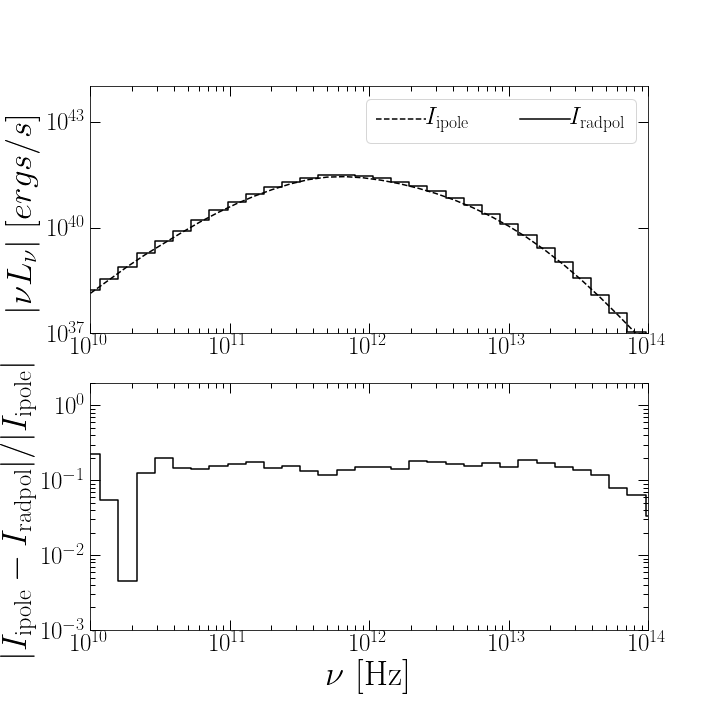}
  \includegraphics[width=0.48\textwidth,angle=0]{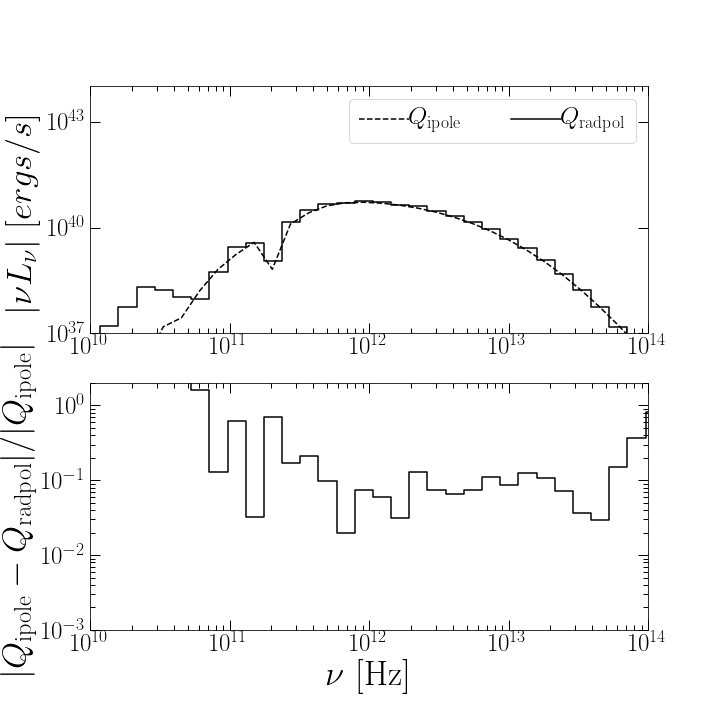}\\
  \includegraphics[width=0.48\textwidth,angle=0]{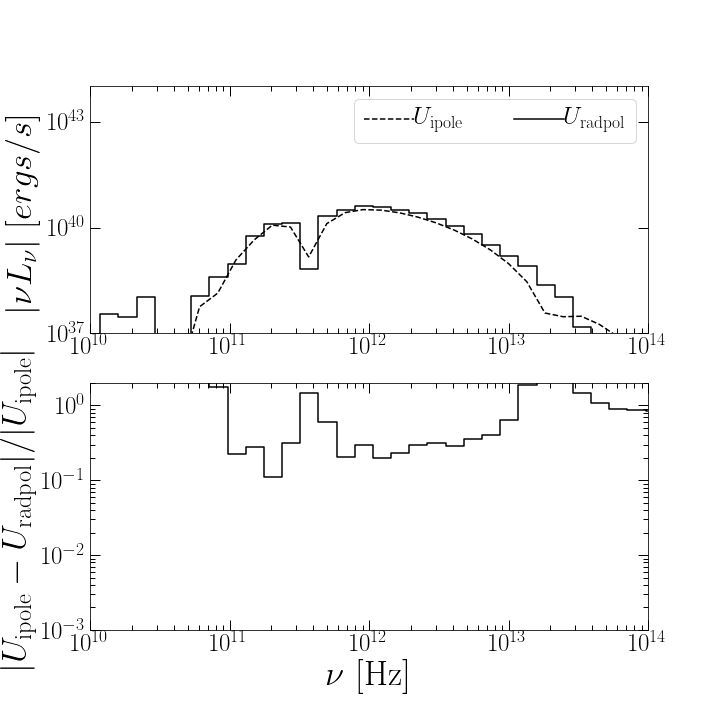}
  \includegraphics[width=0.48\textwidth,angle=0]{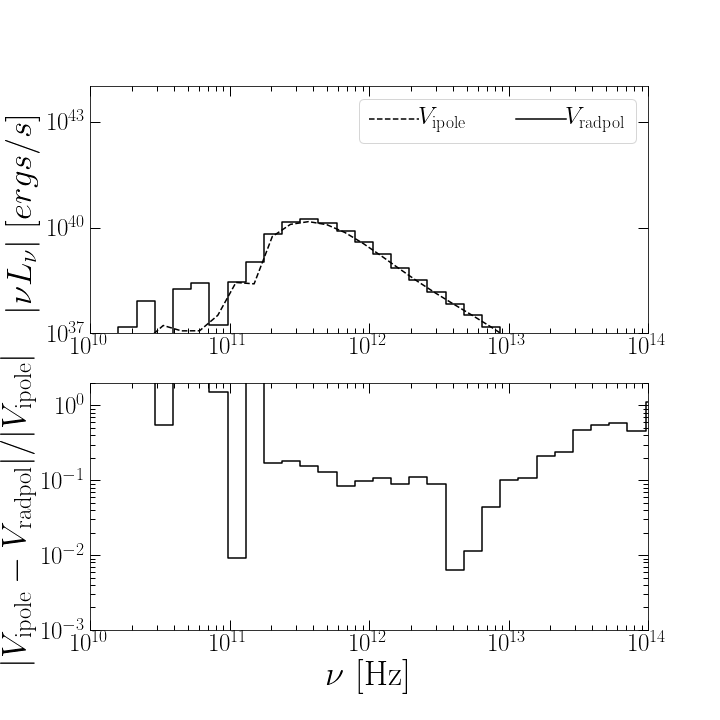}
  \caption{Comparison of the synchrotron spectrum computed 
    with {\tt ipole} ($IQUV_{\rm ipole}$) and {\tt radpol} ($IQUV_{\rm
      radpol}$). Codes agree well in Stokes I, the differences is at the level
    of $\sim 10$ per cent. If the fractional linear or circular
    polarization becomes less than 10 per cent the difference in Stokes QUV becomes
    large.}\label{fig:sed_ipole}
\end{center}
\end{figure*}

In Fig.~\ref{fig:sed_ipole}, we show synchrotron spectrum in all Stokes I, Q,
U and V produced by both codes. The figure shows that results from {\tt
  radpol} converge to those from {\tt ipole}. The agreement between the codes is at the level of 10 per cent in
Stokes I, which is reasonable given the second order integration scheme
implemented in both codes. Codes show
less agreement in low frequency, self-absorbed portion of the synchrotron
spectrum where the fractional polarizations are much lower (due to
selfabsorption and Faraday rotation effects) than the errorbars on Stokes I.
We expect that the differences decrease for smaller step size on
the geodesics.

\subsection{Comparison of SED to original {\tt grmonty} scheme}

\begin{figure*}
\centering
\begin{center}
  \includegraphics[width=0.9\textwidth,angle=0]{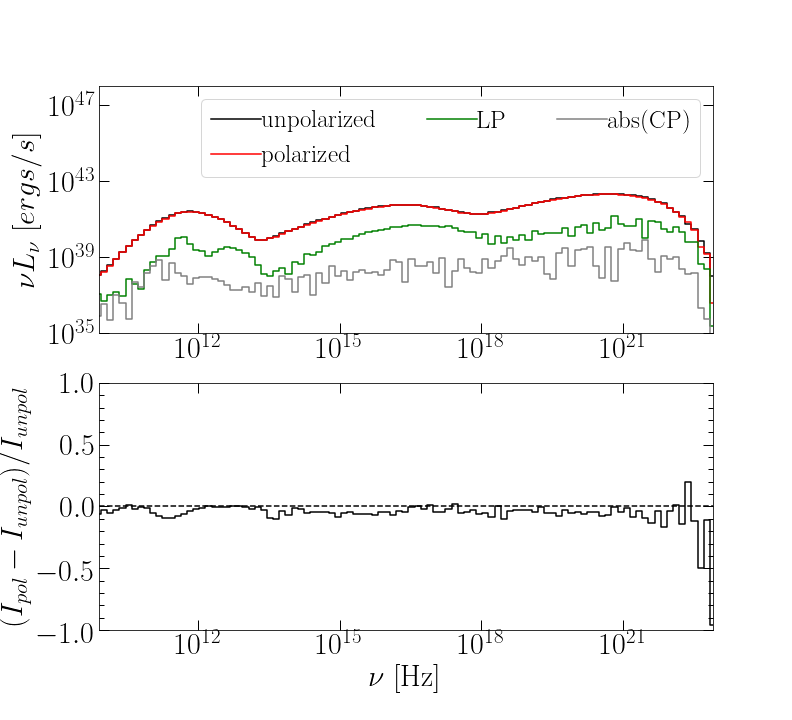}
 \caption{Comparison of the total intensity synchrotron and Compton 
   spectrum emitted by a hot accretion flow around a supermassvie black
   hole. Unpolarized spectra are computed
   using scalar transport in {\tt grmonty} code and the new fully polarized
   spectra are from the presented {\tt radpol} scheme. For the chosen
   parameter of the integration (step size on geodesics) the relative
   difference between the two distinct schemes is less than 10 per cent for most
   of the wavelengths. The largest difference is measured at very high
   energies, near the exponential cut-off of the SED.
   Here we also present the linearly and circularly
   polarized flux of the model.}\label{fig:sed_full}
\end{center}
\end{figure*}

Here we calculate broadband SED (synchrotron+Compton) emitted by a model of a
accretion disk model described in Sect.~\ref{sec:test_setup}. 
In Fig.~\ref{fig:sed_full}, we show comparison of
total intensities of synchrotron emission from the relativistic accretion flow
using original unpolarized radiative transfer scheme {\tt grmonty} and our new
fully polarized radiative transfer scheme. The two spectra agree with each
other within 10 per cent error and the best agreement, within 1 per cent, is in the
optically thin part of the synchrotron spectrum.  The differences decrease
with the size of the step on geodesics. We conclude that our new numerical
scheme {\tt radpol} produces spectra that are consistent with those from {\tt
  grmonty} scheme.

In Fig.~\ref{fig:sed_full} along with the total intensity emission we also
show the expected total linearly and circularly polarized emission from the
hot accretion flow. The circular polarization is low in the present model and
therefore it is quite noisy.  The total linear polarization of the SED is
less than 10 per cent except at very high energy emission at $\nu > 10^{22}$ Hz
where the signal-to-noise decreases.

\section{Summary}\label{sec:conclusions}

We developed a Monte Carlo general relativistic radiative transfer scheme {\tt radpol}. The new scheme
is capable of tracking polarization in the radiative transfer solutions, including
synchrotron emission/absorption, Faraday effects and relativistic Compton scatterings. 
We have tested the new polarimetric scattering kernel against semi-analytic solutions and we compared 
the spectra emitted by a relativistic accretion disk model 
to spectra produced by other radiative transfer codes based on the same model.  
We find that the new code produces polarized synchrotron spectra that converge to those 
from {\tt ipole} code. However the signal-to-noise of the Monte Carlo simulations should be improved 
to reach a better agreement in optically thick part of the synchrotron hump.  

In this work, we consider scattering only on electrons with thermal
relativistic (Maxwell-J{\"u}ttner) energy distribution function. Additional
physics such as synchrotron emission and synchrotron self-Compton scatterings
on non-thermal electron distributions described by a power-law or $\kappa$
functions \citep{livadiotis:2013} can be now easily incorporated and tested.

The new scheme is fully covariant and it has several possible
applications. It can be used to model spectra of relativistic systems such as
gamma-ray bursts jets, X-ray binary systems or active galactic nuclei cores
and jets for which polarization information is available at low and high
energies. The scheme can be also used to estimate under what circumstances
the Compton scattered radiation can be polarized. Such estimates can be now done for
highly non-trivial plasma and magnetic field configurations.

Finally, since the total intensity and the geometry of the high energy radiation field depend on
the polarization state of the seed photons certain high energy processes
should be discussed in context of the current models. 
These processes, that are not included in the present work but ought to
be discussed in a follow-up work,
include: electron-positron pair productions (e.g. \citealt{moscibrodzka:2011}), 
double Compton process \citep{gould:1984}, and stimulated Compton emission
(e.g., \citealt{reinisch:1976}).


\section*{Acknowledgements}
I would like to thank Charles Gammie for his constructive and thoughtful comments on this work.
I also would like to thank Ryuichi Kurosawa for discussions of Monte Carlo methods.
This research has made use of NASA's Astrophysics Data System Bibliographic Services.

\bibliographystyle{mnras}
\bibliography{local}

\begin{figure*}
\centering
\begin{center}
\includegraphics[width=0.48\textwidth,angle=0]{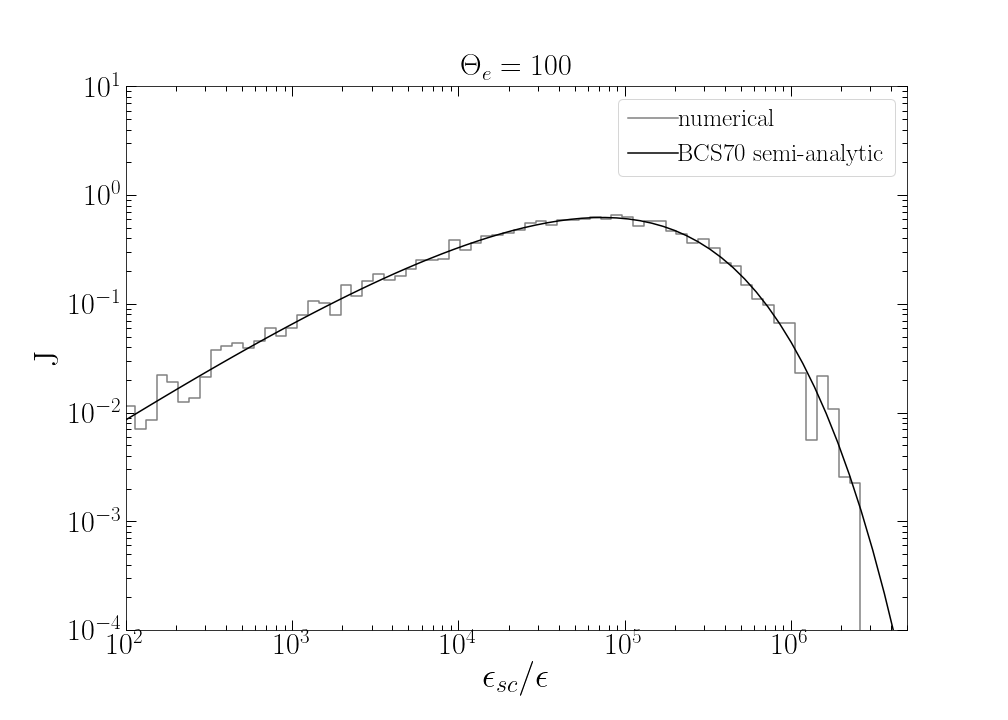}
\includegraphics[width=0.48\textwidth,angle=0]{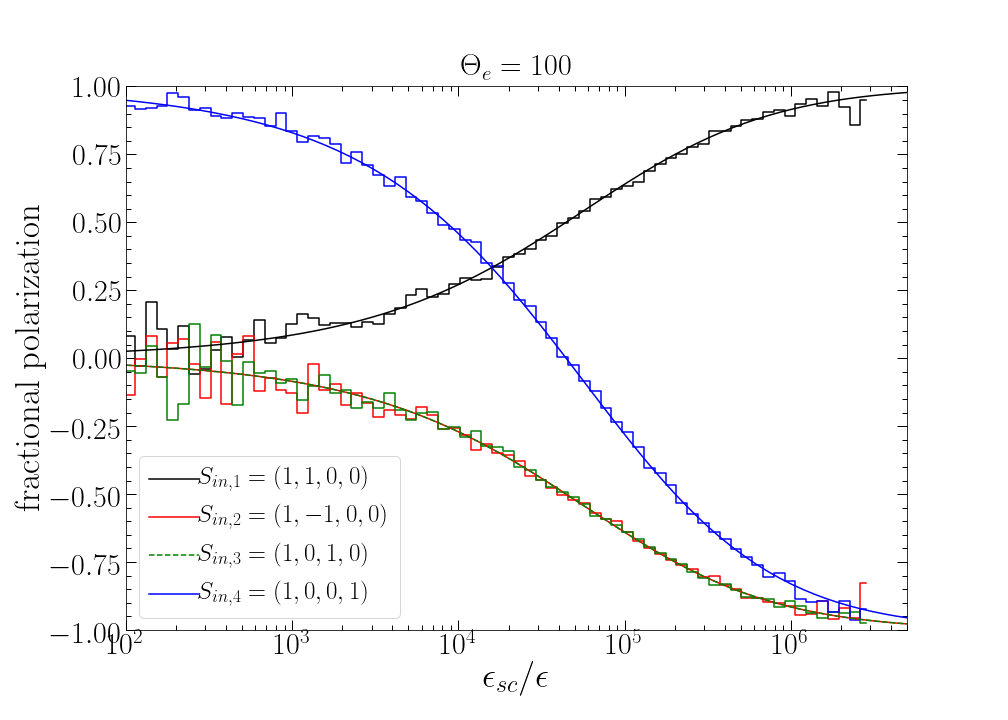}
\caption{Analytical expectation from \citet{bonometto:1970} (lines) and
  numerical results (histograms) for intensity (left panel) and fractional
  polarizations (right panel) of unidirectional, monochromatic light scattered off thermal
  relativistic electrons parameterized with $\Theta_e=100$
  into angle ($\theta',\phi')=(85^\circ,0)$.
  The initial beam energy in the 'fluid' frame is
  $\epsilon=2.5\times 10^{-11}$ to ensure that the scatterings in electron frame
  are elastic.
  We consider four cases when incident beam has
  different polarization state, $S_{in}$. The scattered light fractional polarizations
  agree with theoretical predictions of \citet{bonometto:1970}.
  In this problem, the scattered light total intensity does not depend on the initial beam
  polarization and agrees with theoretical predictions.}\label{fig:bonometto}
\end{center}
\end{figure*}

\appendix

\section{Scattering of a photon beam on an arbitrary distribution of
  electrons: semi-analytic  solution}\label{sec:app}

\citet{bonometto:1970} (see also \citealt{bonometto:1973} and
\citealt{bonometto:1973b})
found a semi-analytic expression for the intensity of a unidirectional and monochromatic photon beam
polarized along $\varepsilon$ scattered by an isotropic distribution of
electrons into $k'$ direction ($\hat{k}'$) into polarization $\varepsilon'$:
\begin{equation}\label{eq:semi-analytic}
J_{\varepsilon'} = K \gamma_{min} \frac{\epsilon'}{\epsilon} \left\{
  \left[\varepsilon \cdot \varepsilon' + \frac{(\hat{k} \cdot \varepsilon')(\hat{k}' \cdot \varepsilon )}{1-\cos \theta'}\right]^2(\Sigma_1+\Sigma_2) + \Sigma_2 \right\}
\end{equation}
where $K=1/2 h c r_0^2 \, {\rm cm^{-3}}$, and
$\gamma_{min}=\sqrt{ \frac{1}{2} \frac{\epsilon'}{\epsilon}
  \frac{1}{1-\cos\theta'} }$ is the minimum Lorentz factor required to produce
scattered photons with energy $\epsilon'$. The quantities $\hat{k}, \hat{k}'$
are directions of incident and scattered photon beam and $\varepsilon$ and
$\varepsilon'$ are their polarizations, measured in the plasma frame. Summing Eq.~\ref{eq:semi-analytic}
over two orthogonal polarizations gives the total intensity of the Compton scattered radiation:
\begin{equation}\label{eq:semi-analytic2}
J_{\varepsilon'} + J_{\varepsilon'_{\bot}}= K \gamma_{min}
\frac{\epsilon'}{\epsilon} (\Sigma_1+3 \Sigma_2).
\end{equation}
The quantities $\Sigma_{12}$ are given in integral forms:
\begin{equation}
  \Sigma_1=\int_0^1 m(\gamma) \left(x^2 - \frac{1}{x^2} +2 \right) dx
\end{equation}
\begin{equation}
  \Sigma_2=\int_0^1 m(\gamma) \frac{(1-x^2)^2}{x^2} dx
\end{equation}
where
\begin{equation}
x=\frac{\gamma_{min}}{\gamma}
\end{equation}
and where $m(\gamma)$ depends on arbitrary chosen isotropic electron distribution function:
\begin{equation}
m(\gamma)=\frac{dn_e}{d\gamma} \gamma^{-2}. 
\end{equation}
\citet{bonometto:1970} show (see also \citealt{krawczynski:2012}) that fractional
polarization of the scattered beam is:
\begin{equation}
\xi_1'=\delta \frac{\Sigma_1+\Sigma_2}{\Sigma_1+3\Sigma_2}.
\end{equation}
where $\delta$ is the fractional polarization of the incident beam with
linear polarization parallel to the scattering plane
(solutions for other incident polarization directions or circular polarization
are also available, see \citealt{bonometto:1970}).
Notice that these semi-analytic solutions are derived only for Thomson scatterings
but are applicable to any electron distribution function. 

Let's consider scattering on Maxwell-J{\"u}ttner distribution of electrons:
\begin{equation}
m(\gamma) = \frac{\gamma \sqrt{ \gamma^2 - 1 } }{K_2(1/\Theta_e)\Theta_e}
e^{-\gamma/\Theta_e} \gamma^{-2}.
\end{equation}
The integrals $\Sigma_{1,2}$ are evaluated using
Gaussian Quadratures in {\it Gnu Scientific Libraries}.
A a photon beam with direction along z-axis
($\hat{k}=(0,0,1)$) has initial energy $\epsilon=2.5 \times 10^{-11}$.
In Fig.~\ref{fig:bonometto} (left panel) we plot the expected semi-analytic (Eq.~\ref{eq:semi-analytic2})
and numerical (produced by our Monte Carlo code) SEDs for electron temperatures $\Theta_e=100$ for photons
scattered in direction described by set of angles ($\theta',\phi'$)=(85$^\circ$,0).
In this test, we considered four cases with different polarizations of incident beam: $S_{in,1}=(1,1,0,0)$,
$S_{in,2}=(1,-1,0,0)$, $S_{in,3}=(1,0,1,0)$, and $S_{in,4}=(1,0,0,1)$ where $S_i$ is the Stokes
parameter. $S_{1,2,3}$ describes a beam that is initially parallel,
perpendicular, 45 angle linear polarization and $S_4$ is circularly polarized.
In Fig.~\ref{fig:bonometto} (right panel) shows scattered photon polarization
in four cases; the numerical results are consisitent with theoretical
expectations.

\bsp
\label{lastpage}
\end{document}